\documentclass[aps,pre,twocolumn,amsmath,amsmath,amssymb,amsfonts,10pt]{revtex4-1}
\usepackage{graphicx}
\usepackage{dcolumn}
\usepackage{bm}
\usepackage{color}
\usepackage{multirow}
\usepackage{amsmath}

\newcommand{\tbf}[1]{\hat{#1}}

\newcommand{\Dis}[1]{\mathcal{D} [#1]\left(\rho\right)}
\newcommand{\am}{\tbf{a}}
\renewcommand{\ap}{\tbf{a}^{\dagger}}
\newcommand{\apt}{(\tbf{a}^{\dagger })^2}
\newcommand{\amt}{\tbf{a}^{2}}
\newcommand{\apth}{(\tbf{a}^{\dagger })^3}
\newcommand{\amth}{\tbf{a}^{3}}
\newcommand{\expect}[1]{\left\langle #1 \right\rangle}
\newcommand{\ket}[1]{| #1 \rangle}
\newcommand{\bra}[1]{\langle #1 |}
\newcommand{\rnmp}[3]{\rho_{#1,#2}^{(#3)}}
\newcommand{\rnmpd}[3]{\dot{\rho}_{#1,#2}^{(#3)}}

\begin{document}
\author{M. R. Jessop, W. Li and A. D. Armour}
\affiliation{Centre for the Mathematics and Theoretical Physics of Quantum Non-Equilibrium Systems and School of Physics and Astronomy, University of Nottingham, Nottingham, NG7 2RD, United Kingdom}
\title{Phase synchronization in coupled bistable oscillators}
\date{\today}

\begin{abstract}
We introduce a simple model system to study synchronization theoretically in quantum oscillators that are not just in limit-cycle states, but rather display a more complex bistable dynamics. Our oscillator model is purely dissipative, with a two-photon gain balanced by single- and three-photon loss processes. When the gain rate is low, loss processes dominate and the oscillator has a very low photon occupation number. In contrast, for large gain rates, the oscillator is driven into a limit-cycle state where photon numbers can become large. The bistability emerges between these limiting cases with a region of coexistence of limit-cycle and low-occupation states.
Although an individual oscillator has no preferred phase, when two of them are coupled together a relative phase preference is generated which can indicate synchronization of the dynamics. We find that the form and strength of the relative phase preference varies widely depending on the dynamical states of the oscillators.
In the limit-cycle regime, the phase distribution is $\pi$-periodic with peaks at $0$ and $\pi$, whilst 
in the low-occupation regime $\pi$-periodic phase distributions can be produced with peaks at $\pi/2$ and $3\pi/2$.
Tuning the coupled system between these two regimes reveals a region where the relative phase distribution has $\pi/2$-periodicity.
\end{abstract}

\maketitle

\section{ Introduction}

The last few years has seen rapid  progress in engineering and probing the properties of nonlinear oscillators in the quantum regime\,\cite{dykman2012,Ong2011,Ong2013,prx2019}. This has stimulated renewed theoretical interest in the properties of such systems, both at the level of individual oscillators and for more complex  many-body realizations\,\cite{Zhang2017,Puri2017,Dykman2018,Rota2019}.
Of particular interest are phenomena, such as synchronization, which result from an interplay between nonlinearity and nonequilibrium features arising from  a combination of gain and loss processes.
In the classical regime, the standard paradigm for synchronization involves a nonlinear oscillator in a limit-cycle state which has a well-defined amplitude, but no preferred phase\,\cite{Pikovsky2003,Acebron2005}. Such oscillators have a tendency to adjust their rhythm to match either a weak external drive or that of other oscillators to which they may be coupled, typically leading to the emergence of a definite phase (or relative phase) for the oscillations. Although synchronization has been studied for  a very long time in classical oscillators\,\cite{Pikovsky2003}, the systematic study of this behavior in quantum oscillators outside the regime where semiclassical approximations work well\,\cite{Cresser1982}  is quite recent\,\cite{Zhirov2006,Manzano2013,Ludwig2013,Mari2013,Lee2013,Walter2014}.

 Studies of synchronization in quantum oscillators have explored issues such as the variety of ways in which the behavior differs from what is found in the semiclassical limit\,\cite{Lee2013,Lorch2016,Davis-Tilley2016,Lorch2017}, how best to quantify the degree of synchronization in the quantum regime\,\cite{Mari2013,Lee2013,Ameri2015,Hush2015}, and the relationship between synchronization and entanglement\,\cite{Mari2013,Manzano2013,Lee2014,Ameri2015,Li2016,Davis-Tilley2016,Galve2017,Roulet2018}.
 Much of the work on quantum synchronization has involved simple models such as the quantum van der Pol (QvdP) oscillator\,\cite{Lee2013,Walter2014,Lorch2016,Lorch2017} and spin-1 particles\,\cite{Roulet2018,Roulet2018a,Koppenhofer2019}, although a range of other systems have also been considered including atomic ensembles\,\cite{Xu2014,Zhu2015} and optomechanical systems\,\cite{Mari2013,Ludwig2013,Weiss2016}. Significant efforts have also been devoted to proposing ways in which the behavior could be probed in experiment using systems such as trapped ions~\cite{Lee2013,Hush2015} or superconducting circuits\,\cite{Nigg2018}.

Although studies of synchronization in the quantum regime have employed a wide range of model systems, they have focussed (with occasional exceptions\,\cite{Zhirov2008}) on systems whose dynamics is essentially just a limit-cycle. Here we instead explore how weak coupling generates synchronization, in the form of particular phase preferences, in a quantum oscillator which has a more complex bistable dynamics. We do this by proposing a minimal model for a quantum oscillator that displays a limit-cycle state as well as a low occupation-number state  (in which the oscillator simply fluctuates about the origin) and can be tuned to a bistable regime in which both of these states coexist.

Our model involves only dissipative processes in which photons (quanta) are lost or gained,  ensuring that an isolated oscillator never has a preferred phase.
The key ingredient of the model responsible for generating bistability is a two-photon gain process.
This is balanced by two channels of photon loss in which either a single photon or three photons are annihilated in the oscillator.
Different dynamical states of the oscillator (low occupation-number regime, limit-cycle and bistability) are achieved by tuning the relative sizes of the gain and loss rates. In many ways the model is a logical extension of the much-studied QvdP oscillator which combines one-photon gain and two-photon loss\,\cite{Lee2013}. However, the QvdP model always displays a limit-cycle, albeit one whose size depends on the ratio of loss and gain rates.

We investigate in detail the phase synchronization that occurs when two of the bistable model oscillators are coupled via a weak photon exchange process.
This leads to a rich range of behavior in the relative phase distribution, with a different pattern of phase preferences emerging depending on the underlying dynamical states of the oscillators. When the gain is strong and the oscillators are in limit-cycle states, maximal values of the relative phase distribution form at $0$ and $\pi$, matching what is usually found for such states(as seen, e.g.\, in the QvdP model\,\cite{Lee2013}).
For oscillators with low occupation numbers and no limit-cycle, however, relative phases of $\pi/2$ and $3\pi/2$ are preferred instead, a result which we argue can be understood as a result of the two-photon gain.
All four peaks emerge simultaneously for a small, intermediate parameter regime. Furthermore, this $\pi/2$-periodic behavior is strongest when the bistability is most pronounced. Interestingly, and in strong contrast to the QvdP oscillator, we find that the strength of the synchronization in the limit-cycle state does not increase with increasing photon numbers.

The outline of the rest of this paper is as follows.
In Sec. \ref{sec:model}, we start by introducing our  bistable quantum oscillator model and exploring its steady-state and dynamical properties.
 Then  we investigate the behavior of two coherently coupled oscillators in Sec. \ref{sec:couple}, focussing in particular on the way in which the pattern and strength of features in the relative phase distribution of the system depend on the underlying dynamical properties of each oscillator. We conclude in Sec. \ref{sec:conc} and the Appendixes provide details about aspects of the calculations employed.

\section{Bistable Oscillator System}\label{sec:model}

\subsection{321-Oscillator Model}
\begin{figure}[t!]
	\includegraphics[width=\linewidth]{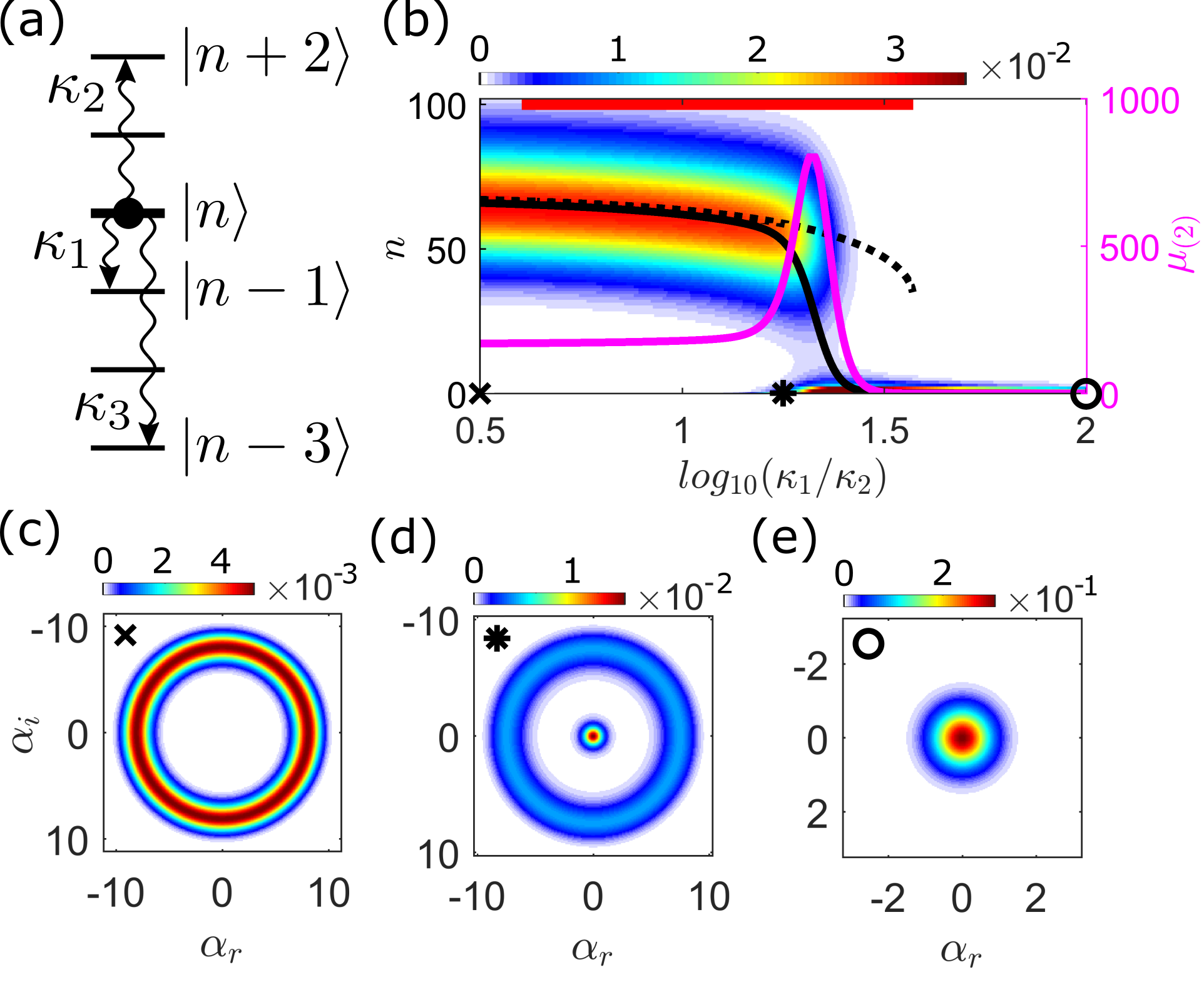}
	\caption{(a) The three dissipative processes of the oscillator: two-photon gain, single-photon loss, and three-photon loss, at rates $\kappa_2,\ \kappa_1$, and $\kappa_3$ respectively. (b) Steady-state properties as a function of $\kappa_1/\kappa_2$ for fixed $\kappa_3=\kappa_2\times10^{-2}$. A color scale shows the photon-number distribution $P_n$ (for $n>0$), with the average photon number $\langle n \rangle$ calculated numerically (solid black line) and mean-field prediction (dotted black line) superposed.
		Large photon number states are occupied when the gain dominates (black cross, $\kappa_2\gg\kappa_1$), the fixed point state is predominantly occupied if the loss dominates (black circle, $\kappa_1\gg\kappa_2$), and a bimodal distribution appears in an intermediate region (black star).
		Also shown are the second moment $\mu_2$ (solid magenta) and the range where the mean-field predicts two stable states (red bar).
		The corresponding Wigner functions, $W\hspace{-3pt}\left(\alpha_r,\alpha_i\right)$, are for (c) limit-cycle ($\kappa_1/\kappa_2=10^{0.5}$), (d) bistability ($\kappa_1/\kappa_2=10^{1.25}$) and (e) fixed point ($\kappa_1/\kappa_2=10^{2}$).}
	\label{fig:1}
\end{figure}

Our oscillator model involves three dissipative processes, as illustrated in Fig. \ref{fig:1}(a). A two-photon gain process with rate $\kappa_2$ drives the oscillator to higher photon numbers, whilst a one-photon loss process damps it at a rate $\kappa_1$; an additional three-photon loss process at rate $\kappa_3$ is included to stabilise the system, ensuring that  it has a steady state for any strength of the gain.
The master equation for a single oscillator in the interaction picture is given by~\cite{Lindblad1976,Gardiner2000,Lee2013},
\begin{equation}\label{eq:ME}
	\dot{\rho}=\mathcal{L}\rho=\kappa_1 \Dis{\am}+\kappa_2 \Dis{\apt}+\kappa_3 \Dis{\amth},
\end{equation}%
where $\hat{a}$ is the oscillator lowering operator and we have defined $\Dis{\hat{C}}=\hat{C}\rho\hat{C}^{\dagger}-\tfrac{1}{2}\{\hat{C}^{\dagger}\hat{C},\rho\}$.
 Our model makes an interesting contrast with the QvdP oscillator~\cite{Lee2013,Walter2014}, where one-photon gain is balanced by two-photon loss. The presence of a non-linear gain process in our model leads to important features, such as bistability, not seen for the QvdP.

The steady-state properties are readily found by exploiting the fact that the system is purely dissipative, so that the dynamics of the diagonal and off-diagonal matrix elements of the density operator in the number (Fock) basis are decoupled~\cite{Scully1997,Scarlatella2018}.
The master equation can therefore be rewritten as a set of $k$ equations
\begin{equation}
\dot{\rho}^{(k)}=\mathcal{M}^{\hspace{-1pt}(k)}\rho^{(k)},\label{eq:M}
\end{equation}
where $\rho_n^{(k)}=\bra{n}\rho\ket{n+k}$, with $|n\rangle$ the $n$-th number state, and $\mathcal{M}^{\hspace{-1pt}(k)}$ a matrix.  
For the diagonal elements,  writing out Eq.\ \eqref{eq:ME} explicitly leads to the coupled set of equations
\begin{equation}
\dot{P}_n = -G_n P_n + A_{n+1} P_{n+1} + B_{n-2} P_{n-2} + C_{n+3} P_{n+3},
\end{equation}%
for the probabilities, $P_n=\langle n|\rho| n\rangle$, with
\begin{align*}
G_n = & \left[ \kappa_1 n + \kappa_2 \left(n+1\right) \left(n+2\right)\right.\left. + \kappa_3 n\left(n-1\right)\left(n-2\right)\right],\\
A_{n+1} = & \kappa_1\left(n+1\right), \\
B_{n-2} =& \kappa_2n\left(n-1\right),\\
C_{n+3} = & \kappa_3\left(n+1\right)\left(n+2\right)\left(n+3\right),
\end{align*}
from which the form of $\mathcal{M}^{\hspace{-1pt}(0)}$ follows.
In the steady state, the off-diagonal terms $\rho^{(k\neq 0)}$ all go to zero and the eigenvector of $\mathcal{M}^{\hspace{-1pt}(0)}$ with zero eigenvalue gives the $P_n$ distribution.

\subsection{Phase diagram}
The steady-state of the oscillator can be characterised by the behavior of the $P_n$ distribution along with the Wigner distribution\,\cite{Gardiner2000}, $W\hspace{-3pt}\left(\alpha_r,\alpha_i\right)$. Figures \ref{fig:1}(b)-(e) show how the state of the system evolves as the ratio $\kappa_1/\kappa_2$ is changed for a small (fixed) value of $\kappa_3/\kappa_2$. When the non-linear gain dominates ($\kappa_1/\kappa_2\ll 1$) the oscillator is driven to large phonon numbers with an almost Gaussian $P_n$ distribution centered at a value $\langle n\rangle=\sum_n nP_n\gg 1$ [see Fig. \ref{fig:1}(b)].
The corresponding Wigner distribution reveals a distribution with a ring of maxima [Fig.\ \ref{fig:1}(c)], we classify this as  a \textit{limit-cycle} (LC) state, as it has a well-defined average amplitude, but no preferred phase\,\cite{Lee2013}.
In the opposite limit of dominant loss ($\kappa_1/\kappa_2\gg 1$), the oscillator is damped to the lowest photon number states, leading to a sharp peak in the $P_n$  distribution at $n=0$. In this regime the Wigner distribution displays a single maximum at the origin [Fig.\ \ref{fig:1}(e)] and we call this a \textit{fixed point} (FP) state.
In between these limits we find \textit{bistability} (B) where features from both the LC and FP states can be found in the Wigner distribution [see Fig.\ \ref{fig:1}(d)] and  two peaks of similar size feature in the $P_n$ distribution\,\cite{rodrigues2007}.

The bimodality of the $P_n$ distribution is captured by a sharp peak in the second moment~\cite{Nielsen2006} $\mu_{(2)}=\langle n^2\rangle -\langle n \rangle ^2$, as shown in Fig. \ref{fig:1}(b).
The distributions with the highest values of $\mu_{(2)}$ are found to be those with the most pronounced bistability, i.e.\ with two peaks of comparable area that are separated by a significant gap.


We use a standard mean-field (or semiclassical) approach to understand the origin of the oscillator's dynamical states (the details are described in Appendix \ref{app:MF1}).
Two physically relevant mean-field solutions for the average photon number are found: a zero photon solution $n_0=0$ (corresponding to the fixed point), and a nonzero solution $n_{+}=\tfrac{\kappa_2}{3\kappa_3}\left[1+\sqrt{1+\tfrac{3\kappa_3}{\kappa_2^2}\left(4\kappa_2-\kappa_1\right)}\right]$ (corresponding to the limit-cycle).
Linear stability analysis reveals that $n_0$ is stable for $\kappa_1>4\kappa_2$, whilst $n_+$ is both non-negative and stable for $3\kappa_1\kappa_3<12\kappa_3\kappa_2+\kappa_2^2$. 
Hence the mean-field approach predicts a region of bistability associated with the coexistence  of these stable solutions in the parameter regime
\begin{equation}
0<\frac{3\kappa_3}{\kappa_2}\left(\frac{\kappa_1}{\kappa_2}-4\right)<1.\label{eq:bicrit1}
\end{equation}
These predictions are compared with the behavior of the $P_n$ distribution  in Fig. \ref{fig:1}(b). The $n_+$ solution matches the average photon number $\langle n\rangle$ very well deep inside the limit-cycle regime, where photon numbers are large. However, the mean-field calculation doesn't describe the extent of the bistable region very accurately, which is not surprising given the strong quantum non-linearity in this regime.

\begin{figure}[t!]
	\includegraphics[width=\linewidth]{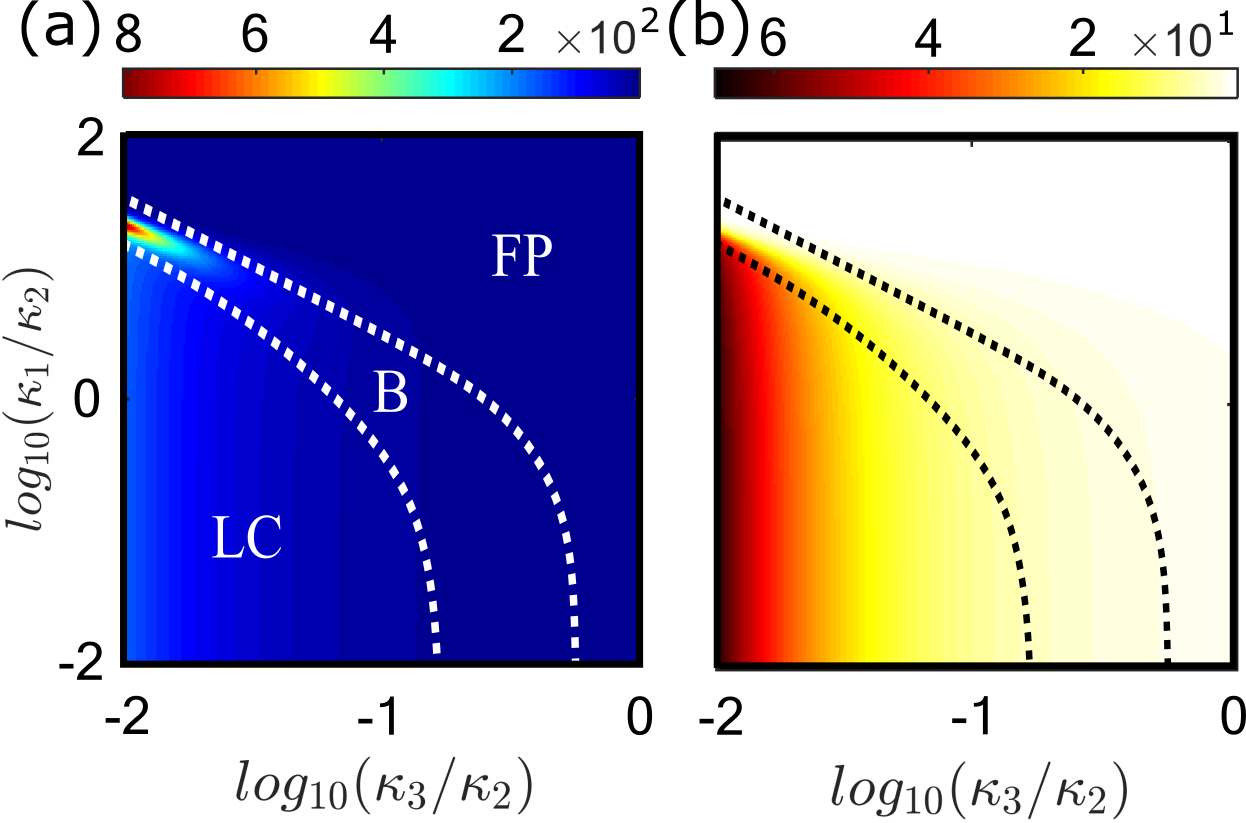}
	\caption{Phase diagrams of (a) the second moment $\mu_{(2)}$  and (b) average occupation number $\langle n\rangle$. Dashed lines indicating the parameter regimes that produce limit-cycle (LC), bistable (B), or fixed point (FP) oscillator states (determined by  analysing the Wigner distributions).
		The  second moment is maximal within the bistable region where the corresponding LC contains a large occupation number (i.e $\kappa_2\gg\kappa_3$), elsewhere it is rather smooth. The  average photon number distribution $\langle n \rangle$ is largest deep within the LC regime (red) and lowest for the fixed point  regions (white), but doesn't provide any indication of bistability.}\label{fig:2}
\end{figure}

The behavior of the oscillator as a function of both $\kappa_1/\kappa_2$ and $\kappa_3/\kappa_2$ is summarised in Fig.\ \ref{fig:2}. Classifying the oscillator state as either FP, B, or LC, based on the corresponding Wigner function leads to the `phase diagram' shown in Fig.\ \ref{fig:2}(a) whilst the behavior of the average occupation number is shown in Fig.\ \ref{fig:2}(b). A sharp transition occurs between the FP and LC states via a bistable region which is narrow and very well defined (e.g.\ via a clear peak in the second moment) for $\kappa_3/\kappa_2\ll 1$ as $\kappa_1/\kappa_2$ is reduced. Average photon numbers tend to decrease as $\kappa_3/\kappa_2$ is increased, and although we can still use the Wigner distribution to distingusih the different states, the differences between them become much less distinct.

\subsection{Dynamical Properties}

We now turn to the dynamical properties of the system to understand whether the bistable states we have identified are also metastable in the sense that they display slow switching between the two states. To investigate this, we start by calculating how the eigenvalues with largest real parts (i.e. least negative, but non-zero) of the system behave\,\cite{risken,Nirmal1996,Macieszczak2016}.

\begin{figure}[t!]
	\includegraphics[width=\linewidth]{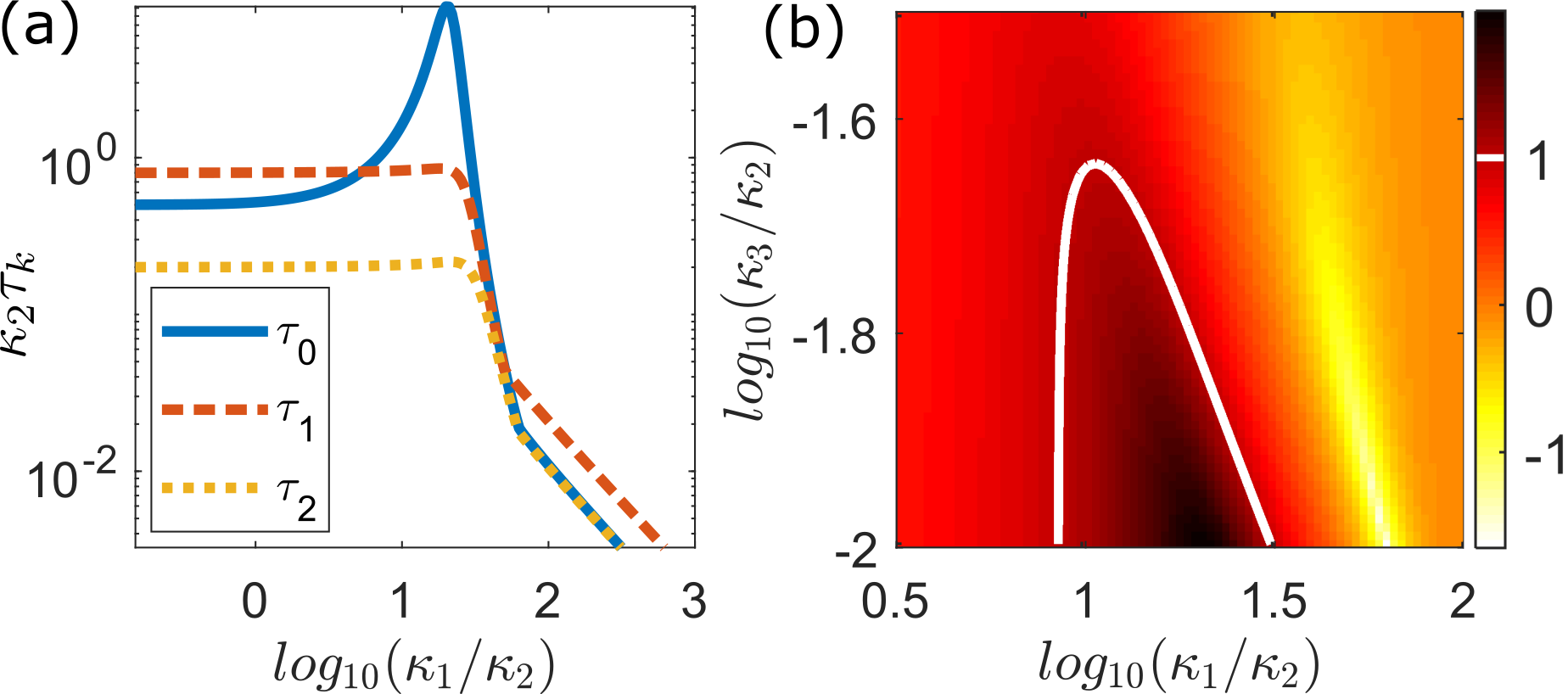}
	\caption{(a) The slowest timescales $\tau_k$ of the oscillator for $k=0,1,2$ with $\kappa_3=\kappa_2\times 10^{-2}$
		and (b) the metastability $M$ [Eq.\ \ref{eq:meta}] plotted on a logarithmic scale. 
		The white contour  is for $M=10$.\label{fig:tau}}
\end{figure}

The largest eigenvalue of the matrix $\mathcal{M}^{\hspace{-1pt}(k)}$ in Eq. \eqref{eq:M} can be used to obtain the slowest timescale associated with the dynamics of $\rho^{(k)}$, $\tau_k=-1/{\rm{Re}}(\lambda^{\hspace{-1pt}(k)}_0)$, examples of which are shown in Fig. \ref{fig:tau}(a). The $k=0$ case, $\tau_0$, describes the relaxation of the diagonal elements and it becomes very large for a range of $\kappa_1/\kappa_2$. The other times-scales $\tau_{1,2}$, describe the relaxation of phase preferences in the system. Although they never become as large as the peak values of $\tau_0$ and display no obvious signature of the bistability, they do change significantly as the system evolves from FP to LC states, becoming orders of magnitude larger in the latter case.

  We can use the emergence of a single very slow timescale as a criterion to identify metastability in the system\,\cite{Macieszczak2016}. We calculate the ratio of the differences between the three largest eigenvalues
  \begin{equation}
  M=\frac{{\rm{Re}}(\lambda^{\hspace{-1pt}(0)}_2-\lambda^{\hspace{-1pt}(0)}_1)}{{\rm{Re}}(\lambda^{\hspace{-1pt}(0)}_1-\lambda^{\hspace{-1pt}(0)}_0)}, \label{eq:meta}
  \end{equation}
   as shown in Fig.\ \ref{fig:tau}(b), and take (somewhat arbitrarily) a separation of $\lambda_1$ and $\lambda_0$ an order of magnitude larger than $\lambda_2$ and $\lambda_1$, i.e.\ $M\geq10$, as a threshold for metastability. The parameter range identified via this criterion is, unsurprisingly, a small subset of that labelled on the basis of the steady-state as bistable in Fig.\ \ref{fig:2}(a). Nevertheless, the peaks in $M$ and the second moment plotted in Fig.\ \ref{fig:2}(a) match up well.

We can get more insight into the dynamical properties of the system by looking at quantum jump trajectories\,\cite{Plenio1998}, obtained by unravelling the master equation.
The system is evolved in time with the non-Hermitian Hamiltonian $H_{MC}=-\frac{i}{2}\left(\kappa_1 \ap\am + \kappa_2 \amt\apt +\kappa_3\apth\amth\right)$.
 For each short time step, quantum jumps of three different kinds (one-photon loss, two-photon gain, or three-photon loss process) can occur with a probability that depends on the state of the system (e.g.\ the probability of two photon gain occurring over the interval $\delta t$ is given by $\kappa_2\langle\psi(t)|\hat{a}^2(\hat{a}^{\dagger})^2|\psi(t)\rangle\delta t$ where $|\psi(t)\rangle$ is the state of the oscillator).

\begin{figure}[t!]
	\includegraphics[width=\linewidth]{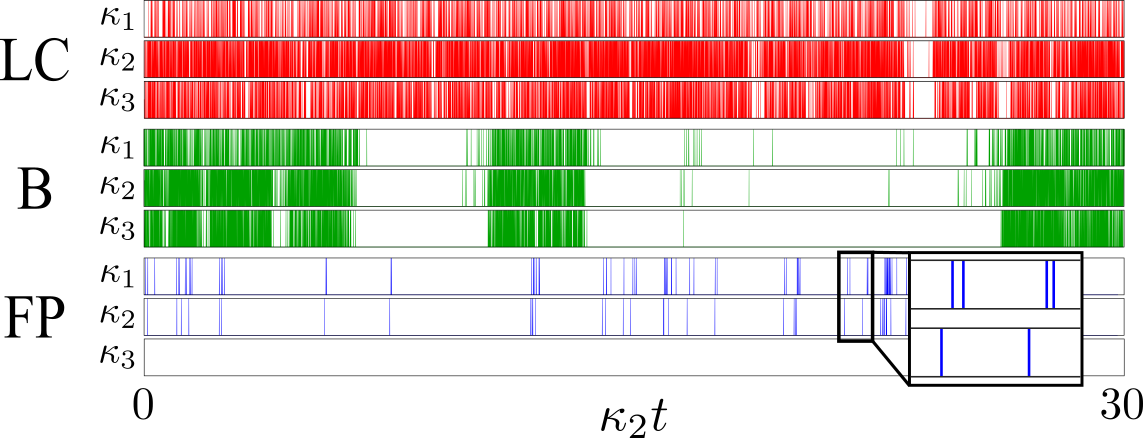}
	\caption{Sample quantum trajectories for each of the oscillator states illustrating the frequency of the different jump processes for (top) limit-cycle ($\kappa_1=\kappa_2\times10^{3/4}$), (middle) bistable ($\kappa_1=\kappa_2\times10^{5/4})$, and (bottom) fixed-point ($\kappa_1=\kappa_2\times10^{7/4}$) states (with $\kappa_3=\kappa_2\times10^{-7/4}$ throughout).  The individual jump processes involving one-photon loss (rate $\kappa_1$), two-photon gain (rate $\kappa_2$), and three-photon loss (rate $\kappa_3$) are indicated. The bistable oscillator can be seen to flip intermittently between LC-like and FP-like behavior. In the FP state the one-photon loss jumps occur in pairs soon after each two-photon gain jump (see magnified portion of the lower panel).  \label{fig:int}}
\end{figure}

The frequency of the different jump processes is illustrated in Fig.\ \ref{fig:int} for sample trajectories obtained for parameters corresponding to the three different states of the system (LC, FP and B). Within the LC state the oscillator has a large number of photons and consequently displays a high level of activity (i.e.\ frequent jumps). In contrast, all jump processes are strongly suppressed (and the three-photon loss especially so) in the FP state because of the very low occupation numbers in this regime.
Within the region which is both bistable and metastable (based on the behavior of the eigenvalues), the oscillator switches between periods in which it exhibits high and low levels of activity; the switching continues indefinitely, never settling into one state or the other.
This intermittency in the dynamics of the trajectories is consistent with our interpretation of this regime as metastable\,\cite{Macieszczak2016}.

\section{Synchronization of Coupled Oscillators}\label{sec:couple}

We now explore how phase ordering and synchronization occurs when two of these oscillators are coupled together weakly. For simplicity, we consider two identical oscillators and assume a coherent (photon-exchange) interaction of the form\,\cite{Lee2013,Walter2014,Hush2015}
\begin{equation}
\hat{H}_J=\hbar J\left(\ap_1\am_2+\am_1\ap_2 \right),
 \end{equation}
 where $\am_j$ is a lowering operator for oscillator $j$ and $J$ is the strength of the coupling. The master equation of the coupled system is given by
\begin{equation}
\dot{\rho} = -\frac{i}{\hbar}\left[\hat{H}_J,\rho\right]+ \sum_{j=1,2}\mathcal{L}_{j}\rho,	\label{eq:MEJ}
\end{equation}
where the dissipation terms follow from Eq.\ \eqref{eq:ME}: $\mathcal{L}_j\rho=\kappa_1 \mathcal{D}[\hat{a}_j](\rho)+\kappa_2 \mathcal{D}[(\hat{a}^{\dagger}_j)^2](\rho)+\kappa_3 \mathcal{D}[\hat{a}^3_j](\rho)$.

\begin{figure}[t!]
	\centering
	\includegraphics[width=\linewidth]{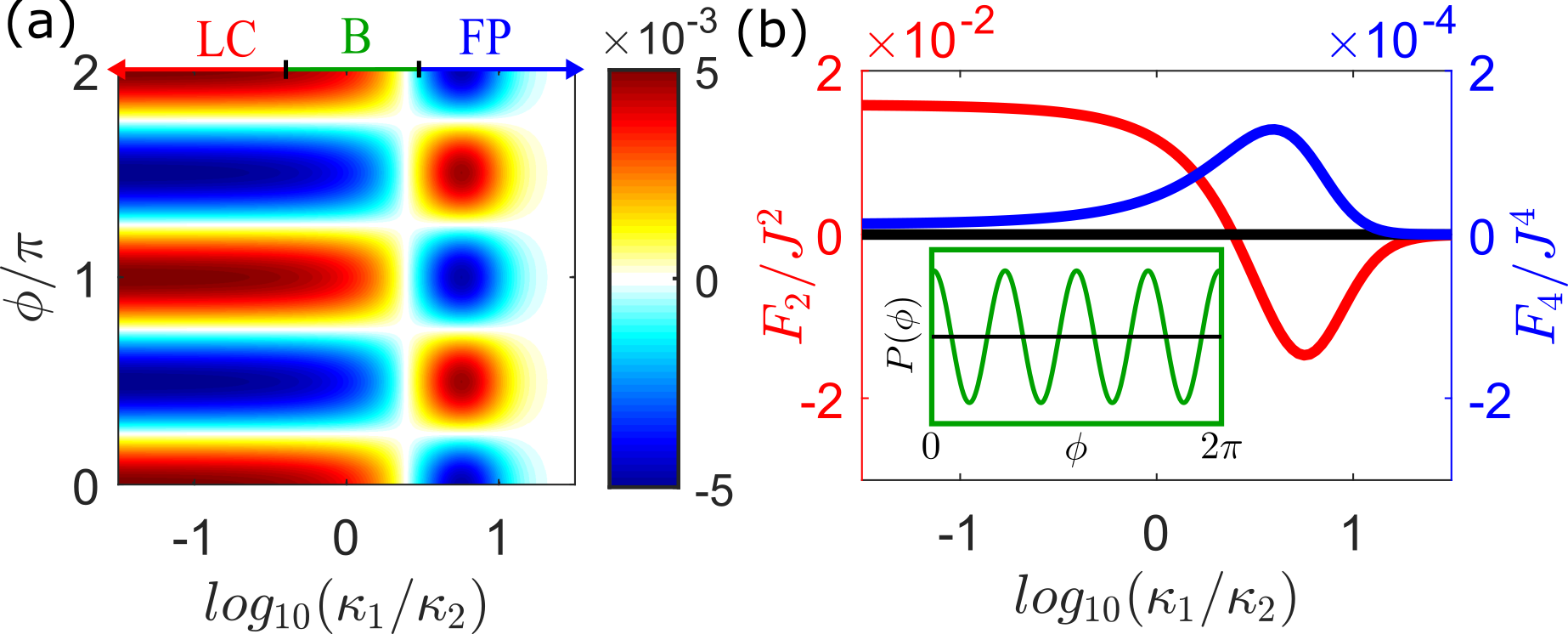}
	\caption{(a) $[P(\phi)-1/2\pi](\kappa_2^2/J^2)$ as a function of $\kappa_1/\kappa_2$ (calculated with perturbation theory), with  $J/\kappa_2=10^{-2}$ and  $\kappa_3=\kappa_2\times10^{-1}$. The range of $\kappa_1/\kappa_2$ shown spans the three states (FP, B, and LC).
		Very weak single-photon loss ($\kappa_1\ll\kappa_2$) leads to coupled limit-cycles with peaks at $0$ and $\pi$.
		As the single-photon loss rate is increased, this $\pi$-periodic pattern vanishes before then reappearing with peaks at $\pi/2$ and $3\pi/2$.
		Eventually, very strong single-photon loss ($\kappa_1\gg\kappa_2$) again suppresses the pattern.
		(b) The dominant Fourier coefficients $F_2$ (red) and $F_4$ (blue) [defined in Eq. \eqref{eq:relph}] (here units are chosen such that $\kappa_2=1$).
		$F_2$ accounts for the $\pi$-periodic component of the relative phase distribution and its sign determines the position of the peaks; $F_4$  is the next largest, though it is much less important than $F_2$, except for the region in which $F_2$ passes through zero. The corresponding $\frac{\pi}{2}$-periodic $P(\phi)$ distribution for the case where $F_2$ is zero is shown in the inset; the peak-to-peak height is $10^{-4}$ in units of $(J/\kappa_2)^4$.\label{fig:Jnum}}%
\end{figure}

Classical limit-cycle oscillators typically synchronize when they are coupled together weakly, developing a preference for one or more relative phase values\,\cite{Pikovsky2003,Kuznetsov2009}.
In the quantum regime, the phase states $|\varphi\rangle=\sum_{n}{\rm{e}}^{i\varphi}|n\rangle/2\pi$, can be used to construct a relative phase probability distribution\,\cite{Barnett1990,Luis1996,Barak2005,Lorch2017,Davis-Tilley2016,Hush2015}
\begin{eqnarray}
P(\phi)&=&\frac{1}{2\pi}\sum_{n,m=0}^{\infty}\sum_{k={\textrm{max}}(n,m)}^{\infty}{\textrm{e}}^{i\phi(m-n)}\nonumber\\
&&\times\langle n,k-n|\rho|m,k-m\rangle,\label{eq:prelp}\\
& = & \frac{1}{2\pi} + \frac{1}{\pi}\text{Re}\hspace{-2pt}\left[\sum_{k=1}^{\infty}e^{ik\phi}\hspace{-4pt}\sum_{n,m=0}^{\infty}\hspace{-4pt}\bra{n+k,m}\rho\ket{n,m+k}\right],\nonumber \\\label{eq:relph}
\end{eqnarray}
where $\phi=\varphi_1-\varphi_2$ is the relative phase of the two oscillators.
If the two oscillators are uncoupled, their phases are independent and the relative phase distribution is uniform, $P(\phi)=\tfrac{1}{2\pi}$.
In the following we use the relative phase distribution to explore the impact of coupling on the behavior of our oscillator model in each of the dynamical states it displays.

\begin{figure}[t!]
	\centering
	\includegraphics[width=\linewidth]{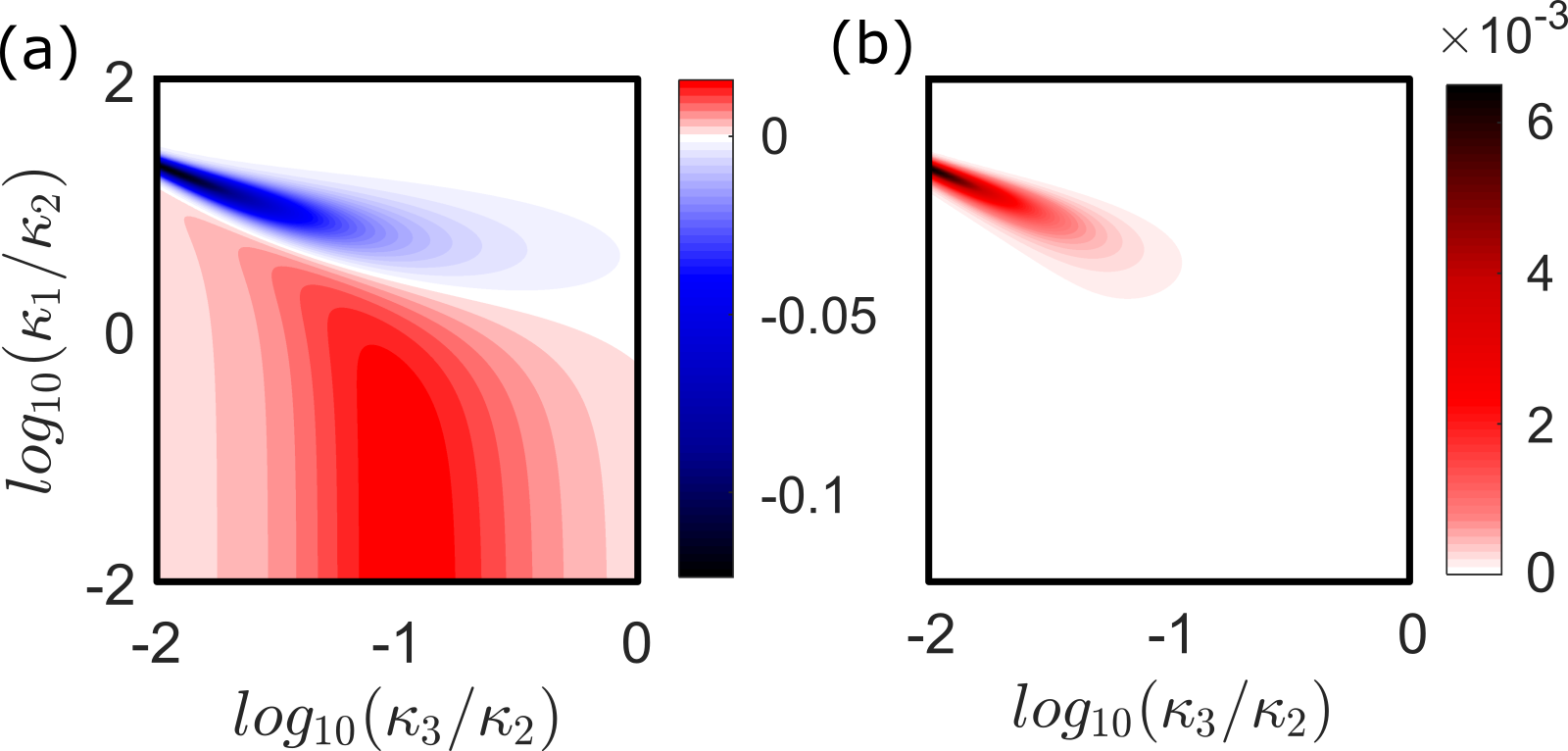}
	\caption{Behavior of (a) $F_2/J^2$ and (b) $F_4/J^4$, the two most dominant Fourier coefficients of the relative phase distribution scaled with coupling strength as a function of the relative loss/gain rates (in units such that $\kappa_2=1)$.\label{fig:f2f4}}%
\end{figure}


Since we are interested in the effect of a very weak coupling between the oscillators, we use a perturbation method\,\cite{Lorch2016,Davis-Tilley2016,Koppenhofer2019} to calculate the relative phase distribution, details of which are described in Appendix  \ref{app:pert}.
Figure \ref{fig:Jnum}(a) shows how the relative phase distribution evolves as a function of $\kappa_1/\kappa_2$ and the system passes from FP to LC via the bistable region. Interestingly, the system displays $\pi$-periodic phase distributions with peaks which reach a similar size in both the LC and FP regimes. However, the location of the peaks are different in the two cases and in between (the bistable region) the phase distribution appears to flatten.

A more detailed picture of the phase behavior is obtained by looking at the Fourier coefficients of the relative phase distribution, $F_k=\text{Re}[\sum_{n,m=0}^{\infty}\bra{n+k,m}\rho\ket{n,m+k}]$. For our coupled oscillator system, these coefficients are non-zero for even $k$ and typically get smaller very rapidly with increasing $k$. The two most important coefficients $F_2$ and $F_4$ are shown in Fig.\ \ref{fig:Jnum}(b) (calculated to second and fourth order in $J$ respectively). The $\pi$-periodicity of the distribution apparent in Fig.\ \ref{fig:Jnum}(a) stems from the fact that the magnitude of $F_2$ is almost everywhere much larger than that of $F_4$. However, as the system transitions from LC to FP (via the bistablity) $F_2$ changes sign to produce the shift in the locations of the peaks; as $F_2$ passes through zero $F_4$ dominates, giving rise to an unusual $\pi/2$ periodic distribution.

The different $\pi$-periodic patterns that arise in the phase distribution can be understood using simple arguments that exploit the specific characteristics of the system in the LC and FP states. Well-within the LC regime photon occupation numbers are large and semiclassical approaches work well as Fig.\ \ref{fig:1}(b) illustrates for the single oscillator case. A straightforward calculation described in Appendix \ref{app:MF3} recovers the preference for relative phases of $0$ and $\pi$ which is generic for coherently coupled limit-cycle oscillators\,\cite{Lee2013,Davis-Tilley2016}.

The pattern of peaks in the relative phase distribution at $\pi/2$ and $3\pi/2$ seen for the FP can be understood by focussing on the limit where $\kappa_1/\kappa_2\gg 1$. In this regime photon occupation numbers are very small, the value of $\kappa_3$ becomes irrelevant, and we can simplify the perturbation theory calculation by assuming only the three lowest Fock states of the oscillators have non-negligible occupations (see Appendix \ref{app:pert} for details). In this limit we find
\begin{equation}
P(\phi)=\frac{1}{2\pi}+J^2f(\kappa_1,\kappa_2)[P_1^2-P_0P_2]\cos(2\phi), \label{eq:pphi2}
\end{equation}
where $f(\kappa_1,\kappa_2)$ is a (positive-valued) function of the two rates, whilst  $P_0$, $P_1$ and $P_2$ are the occupation probabilities of the three lowest Fock states for the corresponding uncoupled oscillator. 
Clearly the behavior is always $\pi$-periodic, but the positions of the maxima depend on the occupation probabilities which in turn depend on the details of how the system is driven and damped in this regime. 
In our oscillator model, the two-photon driving gives a boost to $P_2$, generating a steady-state with a  photon distribution where $P_2P_0/P_1^2> 1$ which leads to the peaks at $\pi/2$ and $3\pi/2$ seen in Fig.\ \ref{fig:Jnum}. 
Thus the fact that our oscillator model displays a $\pi$-periodic relative phase distribution with peaks at $\pi/2$ and $3\pi/2$ can be seen as being due to the particular form of the lowest order nonlinearity in the system, two-photon gain, which shapes the steady-state number distribution.

Applying a very similar analysis to oscillators with different gain and loss processes casts light on the different ways relative phase preferences can develop (within the same low-occupation number limit). In particular, for an oscillator coupled to a thermal bath [with one-photon gain {\emph{and}} loss related by a ratio of rates $\overline{n}/(\overline{n}+1)$, with thermal occupation number $\overline{n}$] and the QvdP oscillator\,\cite{Lee2013} (one-photon gain and two-photon loss) one obtains expressions like that in Eq.\ \eqref{eq:pphi2}. In each case $f$ takes a form that is different in detail,  but the sign of  $P_1^2-P_0P_2$ still determines the location of the peaks. For the QvdP oscillator in the strongly damped limit $P_2P_0<P_1^2$, hence one finds the same phase behavior as in the large photon number regime\,\cite{Lee2013} (peaks at $0$ and $\pi$). In contrast, for a thermal oscillator $P_2P_0=P_1^2$ and so no phase preference is expected.

Finally, we look at how the phase behavior of the system varies over the full phase diagram. Figure \ref{fig:f2f4} shows how the components $F_2$ and $F_4$ behave for a range of $\kappa_3/\kappa_2$ as well as $\kappa_1/\kappa_2$. The transition between negative and positive values of $F_2$ is sharpest, and the corresponding peak in $F_4$ strongest, in the region where $\kappa_3/\kappa_2\ll1$, which is where the system also displays metastability.

Figure  \ref{fig:f2f4} also reveals a somewhat surprising feature of the phase behavior. The strongest phase preferences are not associated with the limit-cycle regime. Furthermore, deep within the limit-cycle regime the magnitude of $F_2$ starts to saturate, even as the average photon number continues to grow, i.e.\ as $\kappa_3/\kappa_2$ is further reduced in the bottom left quadrant of Fig.\ \ref{fig:f2f4}(a). This is in contrast to other quantum limit-cycle oscillators, such as the QvdP oscillator, for which phase synchronization effects are enhanced by increasing the photon number\,\cite{Lee2013}. We can gain some insight into why this occurs from the phase dynamics of an uncoupled oscillator. At least in the semiclassical regime of large photon numbers, we expect to see stronger synchronization effects emerge when coupling is introduced in systems where the phase diffusion is weaker\,\cite{Pikovsky2003,Davis-Tilley2016}.

Using an approximate analytic approach (see Appendix \ref{app:phasdiff} for details), we find that the phase diffusion rate is simply proportional to $\kappa_2$ in the regime where gain dominates over losses and photon numbers are large. This contrasts with the behavior of the  QvdP oscillator where the phase diffusion is  $\propto 1/\langle n\rangle$, which is why for this system synchronization effects get stronger as photon numbers are increased.

\section{Conclusions} \label{sec:conc}
We have introduced a simple oscillator model with a two-photon gain process balanced by one- and three-photon losses that can be used to engineer a bistable oscillator state. The bistability occurs when the gain/loss rates are tuned between regimes where the oscillator displays a limit-cycle state characterised by a large amplitude (but no preferred phase) and a fixed-point state where occupation numbers are low. Quantum trajectory simulations show clear evidence of metastability in the bistable region, signalled by intermittency in the frequency of the quantum jumps.

When two such oscillators are coupled together weakly their relative phase distribution displays a  rich pattern of behavior. Whilst the system has the usual predominantly $\pi$-periodic distribution with peaks at $0$ and $\pi$ within the limit-cycle regime, in the limit of low occupation numbers the peaks appear at $\pi/2$ and $3\pi/2$. In between these two regimes, where the bistability arises, the distribution can instead be $\pi/2$-periodic.
We identify the presence of two-photon gain as the key factor giving rise to this unusual behavior.

Our goal in this work has been to explore the properties of the relative phase distribution and its connection to the underlying oscillator dynamics in the simplest possible  model system displaying limit-cycles and bistability. We plan to address the question of how the complex behavior we uncovered could be realised, and detected, in future work.
  It would also be interesting to investigate how the dynamics is affected by strong couplings\,\cite{Ishibashi} and the patterns of phase behavior that arise in systems where more than two oscillators are coupled together\,\cite{Ludwig2013,Lee2013,DT}.

\section*{Acknowledgements}
This work was supported by the Engineering and Physical Sciences Research Council [grant number EP/N50970X/1] through a studentship held by MJ.
W.L. acknowledges support from the UKIERI-UGC The- matic Partnership No. IND/CONT/G/16-17/73, EP- SRC Grant No. EP/M014266/1 and EP/R04340X/1.
\appendix


\section{Mean-Field Analysis}\label{app:MF1}
An equation of motion for the expectation value of the annihilation operator, $\frac{d}{dt}\hspace{-2pt}\expect{\am}=\text{Tr}\hspace{-2pt}\left[\am\dot{\rho}\right]$, is found from the master equation [Eq. \eqref{eq:ME}]
\begin{equation}
\frac{d}{dt}\expect{\am} = -\frac{\kappa_1}{2}\expect{\am}+\kappa_2\left(2\expect{\am}+\expect{\ap\amt}\right)-\frac{3\kappa_3}{2}\expect{\apt\amth}.\label{eq:adot}
\end{equation}
In a mean-field (or semiclassical)  approach we break the correlations between operators such that $\langle \hat{A} \hat{B}\rangle=\langle \hat{A}\rangle\langle \hat{B}\rangle$.
We choose to carry out this approximation after first normal ordering the operators.

Making the substitution $\langle a\rangle=r{\rm{e}}^{i\varphi}$ where $r$ and $\varphi$ are classical amplitude and phase variables, allows us to rewrite Eq. \eqref{eq:adot} as
\begin{equation}
\dot{r}+ir\dot{\varphi} = -\frac{3\kappa_3}{2}r^5+\kappa_2r^3+\left(2\kappa_2-\frac{\kappa_1}{2}\right)r.\label{eq:mf1rdot}
\end{equation}
Evaluating the real and imaginary parts separately, leads to $\dot{\varphi}=0$ and
\begin{equation}
\dot{r}=\left[-\frac{3\kappa_3}{2}r^4+\kappa_2r^2+\left(2\kappa_2-\frac{\kappa_1}{2}\right)\right]r,\label{eq:MFrdot}
\end{equation}
with steady state solutions $n_0=0$ and
\begin{equation}
n_{\pm}=\frac{\kappa_2}{3\kappa_3}\left[1\pm\sqrt{1+\frac{3\kappa_3\left(4\kappa_2-\kappa_1\right)}{\kappa_2^2}}\right]. \label{eq:npm}
\end{equation}
These solutions correspond to the average photon number and so are subject to the constraints of being both positive and real.
The stability of the solutions can be determined, e.g.\ through the properties of the relevant  Jacobian.

The negative branch, $n_-$, is never both physical and stable, the zero photon solution, $n_0$, is stable for $\frac{\kappa_1}{\kappa_2} > 4$, and the positive branch, $n_+$, is stable whenever it is physical, i.e.\ for $\frac{\kappa_1}{\kappa_2}<4+\frac{\kappa_2}{3\kappa_3}$.
This mean-field calculation results in two stable solutions for the average photon number and therefore a predicted bistability in the photon number for the parameter regime
\begin{equation}
0<\frac{3\kappa_3}{\kappa_2}\left(\frac{\kappa_1}{\kappa_2}-4\right)<1.
\end{equation}

\section{Perturbation theory}\label{app:pert}
\subsection{General Method}
Perturbation theory provides a convenient way of calculating the way in which the relative phase distribution behaves for weak coupling.
The steady-state of the uncoupled ($J=0$), two-oscillator system [Eq. \eqref{eq:MEJ}] only has diagonal terms.
Treating the coupling as a perturbation\,\cite{Lorch2017,Koppenhofer2019} allows us to calculate the terms in the first off-diagonal as a function of the uncoupled oscillator terms.
Each subsequent off-diagonal can, in turn, be calculated from previous ones\,\cite{Davis-Tilley2016}.

Writing  Eq. \eqref{eq:MEJ} in the number state basis, with $\rnmp{n}{m}{p}=\bra{n+p,m}\rho\ket{n,m+p}$, leads to a set of simultaneous equations
\begin{align}
\rnmpd{n}{m}{p}=& +i J \Delta_{n,m}^{(p)} -\left[ G_{n}^{(p)}+G_{m}^{(p)}\right]\rnmp{n}{m}{p} \notag\\
&+ A_{n+1}^{(p)}\rnmp{n+1}{m}{p}+ B_{n-2}^{(p)}\rnmp{n-2}{m}{p}+ C_{n+3}^{(p)}\rnmp{n+3}{m}{p}\notag\\
&+ A_{m+1}^{(p)}\rnmp{n}{m+1}{p}+ B_{m-2}^{(p)}\rnmp{n}{m-2}{p}+ C_{m+3}^{(p)}\rnmp{n}{m+3}{p},\label{eq:rnmp}
\end{align}
with
\begin{widetext}
\begin{align}
\Delta_{n,m}^{(p)} = &-\sqrt{(n+1)(m+p)}\rnmp{n+1}{m}{p-1}+\sqrt{(m+1)(n+p)}\rnmp{n}{m+1}{p-1}
-\sqrt{n(m+p+1)}\rnmp{n-1}{m}{p+1}+\sqrt{m(n+p+1)}\rnmp{n}{m-1}{p+1}\label{eq:Delta},\\
G_{n}^{(p)} =& \frac{1}{2}\left\{\kappa_1(2n+p)+\kappa_2[(n+p+1)(n+p+2)+(n+1)(n+2)]\right.\notag\\
&\left.\qquad+\kappa_3[(n+p)(n+p-1)(n+p-2)+n(n-1)(n-2)]\right\},\\
A_{n+1}^{(p)} =& \kappa_1\sqrt{(n+1)(n+p+1)},\\
B_{n-2}^{(p)} =& \kappa_2\sqrt{n(n-1)(n+p)(n+p-1)},\\
C_{n+3}^{(p)}=&\kappa_3\sqrt{(n+1)(n+2)(n+3)(n+p+1)(n+p+2)(n+p+3)}.
\end{align}
\end{widetext}
In the steady-state, this reduces to sets of simultaneous equations with the coupling term, $\Delta^{(p)}_{n,m}$ coupling together terms with different $p$-values.
The zeroth-order terms are the diagonal ($p=0$) elements, the uncoupled probabilities $\rnmp{n}{m}{0}=P_nP_m$, which necessarily sum to unity.
The first-order terms are obtained by substituting the zeroth-order terms into the expression for $\Delta_{n,m}^{(p)}$, leading to non-zero contributions for $p=1$ and the process is continued to higher order in $J$ recursively.

The first-order terms obey the relation $\rnmp{m}{n}{1}=-\rnmp{n}{m}{1}$ and hence sum to zero\,\cite{Davis-Tilley2016}, which means that they make  no contribution  to the relative phase distribution [Eq.\ \eqref{eq:relph}] since it depends on {\emph{sums}} of the off-diagonal elements.
The sum of the $p=2$ terms, however, is real and finite and so does contribute resulting in a $\pi$-periodic relative phase distribution.
Continuing to higher orders, we find that all of the odd-$p$ terms sum to zero, and so only the even $p$ sums contribute to the relative phase distribution.
In particular, the $p=4$ terms, lead to  a ${\pi}/{2}$-periodic contribution which can dominate the phase distribution when the $\pi$-periodic terms vanish.

\subsection{Low Occupation-Number Regime}
This calculation can be simplified and solved analytically in the limit of very low photon numbers.
We proceed by assuming only the lowest three photon states are appreciably occupied, i.e. $P_{n>2}=0$, and hence truncate the state space to include only $\ket{0}$, $\ket{1}$, and $\ket{2}$.
Due to the size of the Hilbert space,  only a single term contributes to the relative phase distribution, $P(\phi) = \frac{1}{2\pi} + \frac{1}{\pi}\text{Re}\hspace{-2pt}\left[e^{2i\phi}\rnmp{0}{0}{2}\right]$.
In the steady-state, Eq.\ \eqref{eq:rnmp} with $p=2$ leads to
\begin{equation}
	\rnmp{0}{0}{2}=-iJ2\sqrt{2}\left(2\kappa_1+14\kappa_2\right)^{-1}\rnmp{1}{0}{1},
\end{equation}
using the relation $\rnmp{m}{n}{1}=-\rnmp{n}{m}{1}$.
Equation \eqref{eq:rnmp} with $p=1$ gives
\begin{equation}
	\rnmp{1}{0}{1}=\frac{iJ\sqrt{2}\left(P_1^2-P_0P_2\right)}{\left(2\kappa_1+13\kappa_2\right)}.
\end{equation}
This results in the relative phase distribution
\begin{equation} P\left(\phi\right)=\frac{1}{2\pi}+\frac{2J^2\left(P_1^2-P_0P_2\right)\cos\left(2\phi\right)}{\pi\left(\kappa_1+7\kappa_2\right)\left(2\kappa_1+13\kappa_2\right)}.
\end{equation}
This is a $\pi$-periodic distribution and the position of the peaks is determined by the steady state of the uncoupled oscillators. The two-photon driving in our model ensures $P_2P_0>P_1^2$, which leads to peaks at $\pi/2$ and $3\pi/2$.


\section{Mean-Field Analysis of Coupled Oscillators}\label{app:MF3}
The same mean-field procedure discussed above in Appendix \ref{app:MF1} can be applied to the case of two coupled oscillators with $\alpha=\expect{\am_1}$ and $\beta=\expect{\am_2}$, leading to the equations of motion
\begin{eqnarray}
\dot\alpha &=& -\frac{\kappa_1}{2}\alpha+\kappa_2\left(2\alpha+\alpha|\alpha|^2\right)-\frac{3\kappa_3}{2}\alpha|\alpha|^4-iJ\beta,\label{eq:alphaJdot}\\
\dot\alpha &=& -\frac{\kappa_1}{2}\beta+\kappa_2\left(2\beta+\beta|\beta|^2\right)-\frac{3\kappa_3}{2}\beta|\beta|^4-iJ\alpha.\label{eq:betaJdot}
\end{eqnarray}
Changing to polar co-ordinates, with the definitions $\alpha=r_1e^{i\phi_1}$ and $\beta=r_2e^{i\phi_2}$, and then rewriting the equations in terms of the sum-and-difference variables $r=r_1-r_2$, $R=r_1+r_2$, and $\phi=\phi_1-\phi_2$ we find
\begin{align}
\dot{\phi}=&\frac{4JrR}{R^2-r^2}\cos\phi,\label{eq:MFJp}\\
\dot{r}=&r\left(2\kappa_2-\frac{\kappa_1}{2}\right)+\frac{\kappa_2}{4}r\left(R^2+r^2\right)\notag\\
&-\frac{3\kappa_3}{32}r\left(5R^4+10R^2r^2+r^4\right)-JR\sin\phi\label{eq:MFJr}.
\end{align}
For large photon occupation numbers (as is the case in the limit-cycle regime) and for weak couplings ($J/\kappa_2\ll1$), we have $r\ll R$ and can approximate these  equations as
\begin{align}
\dot{\phi}\approx&\frac{4Jr}{R}\cos\phi,\label{eq:MFJp2}\\
\dot{r}\approx&-\frac{15\kappa_3R^4}{32}r-JR\sin\phi\label{eq:MFJr2}.
\end{align}
The weakness of the coupling and the small size of the ratio $r/R$ leads to a separation of timescales with $\phi$ relaxing much more slowly than $r$. Adiabatic elimination of $r$ leads to the simple relation $\dot{\phi}=-{\partial U(\phi)}/{\partial \phi}$  with the pseudo-potential\,\cite{Kuznetsov2009}
\begin{equation}
	U(\phi)=-\frac{16J^2}{15\kappa_3 R^4}\cos 2\phi.
\end{equation}
This potential predicts relative phase preferences for $0$ and $\pi$, its stable minima.


\section{Phase Diffusion}\label{app:phasdiff}
In this Appendix we return to the case of a single oscillator and obtain an estimate for the phase diffusion rate in the semiclassical limit where photon numbers are large. In the semiclassical regime at least, the strength of phase synchronization in coupled oscillators is determined by competition between the coupling and the rate of phase diffusion in the individual oscillators, with slower phase diffusion leading to stronger phase preferences\,\cite{Davis-Tilley2016,Pikovsky2003}

The phase distribution for a single oscillator takes the form\,\cite{Knight2004,Barnett1990,Davis-Tilley2016}
\begin{eqnarray}
P(\varphi)&=&\frac{1}{2\pi}\sum_{n,m=0}^{\infty}\langle n|\rho|m\rangle{\rm {e}}^{i(m-n)\varphi},\\
&=&\frac{1}{2\pi}+\frac{1}{\pi}{\rm{Re}}\left[\sum_{k=1}^{\infty}{\rm {e}}^{ik\varphi}\Phi^{(k)}\right], \label{eq:defn}
\end{eqnarray}
with $\Phi^{(k)}=\sum_{n=0}^{\infty} \rho^{(k)}_n$, where $\rho^{(k)}_n=\bra{n}\rho\ket{n+k}$.
Although the behavior is in general quite complex, we can obtain a simple approximate description in the semiclassical limit where the density matrix is tightly peaked around a large average photon occupation number\,\cite{Scully1997,Davis-Tilley2016}.

Using \eqref{eq:ME} [and the notation introduced in \eqref{eq:rnmp}], we can obtain the equation of motion for $\Phi^{(k)}$
\begin{equation}
\dot{\Phi}^{(k)}=\kappa_2\sum_n\left[-G_n^{(k)}+A_n^{(k)}+B_n^{(k)}+C_n^{(k)}\right]\rho_{n}^{(k)}.
\end{equation}
In the semiclassical limit, i.e.\ the strong gain regime where $\gamma=\kappa_1/\kappa_2\ll1$ and $\Gamma=\kappa_3/\kappa_2\ll1$, the  photon number saturates to a large value $\langle n\rangle\simeq2\kappa_2/(3\kappa_3)$ [see Eq. \eqref{eq:npm}]. We proceed by assuming we can replace $n$ by $\langle n\rangle$, and expand the square-roots appearing in $A_n^{(k)}$, $B_n^{(k)}$, etc, treating $1/\langle n\rangle$ together with $\gamma$ and $\Gamma$ as small quantities\,\cite{Scully1997,Davis-Tilley2016,Schieve}.
This leads to the simplified equation
\begin{equation}
	\dot{\Phi}^{(k)}=\kappa_2\left[ -\tfrac{5}{4}k^2 + \mathcal{O}\left(\gamma,\Gamma,\langle n \rangle^{-1}\right) \right]\Phi^{(k)}.\label{eq:phidot}
\end{equation}
Hence to leading order the relaxation timescale for the $k$-th component, $\Phi^{(k)}$,  is simply proportional to $1/\kappa_2$, $\tau_k^{LC}\kappa_2\simeq 4/(5k^2)$.

Notice that the semiclassical approximation here assumes all the off-diagonal elements ($\rho_{n}^{(k)}$) decay at the same rate, hence the decay time for $k=1$ is also an approximate linewidth for the oscillator\,\cite{Scully1997}.
The slowest timescales $\tau_k$ associated with the matrices $\mathcal{M}^{\hspace{-1pt}(k)}$  in Eq. \eqref{eq:M} are found to plateau in the limit of $\kappa_1/\kappa_2\ll1$ and $\kappa_3/\kappa_2\ll1$ [illustrated on the left-hand side of Fig. \ref{fig:tau}(a)] and numerically we find $\tau_1\simeq0.8/\kappa_2$ and $\tau_2\simeq0.2/\kappa_2$ in this regime, matching up very well with $\tau_1^{LC}$ and $\tau_2^{LC}$, respectively.

Finally, using the definition Eq.\ \ref{eq:defn} and Eq.\ \ref{eq:phidot}, we see that the phase distribution obeys a diffusion equation\,\cite{Davis-Tilley2016,Schieve}
\begin{equation}
\dot{P}({\varphi})=\frac{5\kappa_2}{4}\frac{\partial^2 P(\varphi)}{\partial \varphi^2}.
\end{equation}
This is very different to what is found obtained in a similar calculation for the QvdP oscillator (or indeed the laser\,\cite{Scully1997}) where the diffusion constant is  $\propto 1/\langle n\rangle$. Hence for such systems phase diffusion gets weaker (and the linewidth narrower), so that synchronization effects get stronger, as the photon number increases\,\cite{Lee2013}.

\end{document}